\documentclass[conference]{IEEEtran}

\ifCLASSINFOpdf
  \usepackage[pdftex]{graphicx}
\else
\fi

\usepackage{amsmath}
\usepackage[colorlinks=true, allcolors=blue]{hyperref}

\begin{document}

\title{Enhancing the physical layer security \\ with bending beams}

\author{\IEEEauthorblockN{Sotiris Droulias*}
\IEEEauthorblockA{Department of Digital Systems\\
University of Piraeus\\
18534 Piraeus, Greece\\
Email: sdroulias@unipi.gr}
\and
\IEEEauthorblockN{Giorgos Stratidakis}
\IEEEauthorblockA{Department of Digital Systems\\
University of Piraeus\\
18534 Piraeus, Greece\\
Email: giostrat@unipi.gr}
\and
\IEEEauthorblockN{Angeliki Alexiou}
\IEEEauthorblockA{Department of Digital Systems\\
University of Piraeus\\
18534 Piraeus, Greece\\
Email: alexiou@unipi.gr}
}

\maketitle

\begin{abstract}
Wavefront engineering for applications in near-field wireless connectivity is gradually becoming common ground. In this landscape, beams that propagate on bent paths are ideal candidates for dynamic blockage avoidance and suppression of potential eavesdropping. In this work we study the physical layer security offered by bending beams, and we demonstrate their capabilities for line-of-sight and non-line-of-sight eavesdropping. We analyze the dependencies between the possible locations of an eavesdropper and the design parameters of such beams, and we introduce metrics to assess their physical layer security performance. Our results demonstrate their superiority with respect to beams generated with conventional beam-forming.
\end{abstract}

\IEEEpeerreviewmaketitle

\section{Introduction}
\noindent As the interest of wireless communications is gradually shifting from the far-field to the near-field of antenna systems, the concept of wavefront engineering for applications in near-field wireless connectivity is gradually becoming common ground. Tailoring the beam wavefront to acquire curvature beyond the typical far-field planar form used in conventional beam-forming, opens up opportunities for generating new types of beams that can focus power at selected areas \cite{Eldar2022, Dai2023, Droulias2024a}, propagate without diffracting \cite{Droulias2024b,Mutai2025} or even evolve along bent paths \cite{Droulias2025,Uchimura2026}.   \\
\indent Beams that propagate on bent paths, or simply \textit{bending beams}, can circumvent undesired objects or users and, owing to this unique property, these beams are ideal for blockage avoidance \cite{Guerboukha2024,Chen2025}. Along these lines, because bending beams can circumvent a potential eavesdropper that attempts to tap the communication link, such beams are expected to provide enhanced physical layer security (PLS) over conventional beamforming. The possibility to increase PLS is still unexplored, with only a few works demonstrating the PLS capabilities of such beams. In \cite{Petrov2024, Petrov2025} the concept of wavefront hopping was recently introduced, in which bending beams (of Airy type) are used in combination of other beam types (Bessel), to maximize the PLS. This is achieved by assigning different parts of a message to different beams, which are used interchangeably in a time domain scheme to communicate with the user, so that the eve misses some parts and is not able to decode the entire message. Only very recently \cite{Stults2026}, the first experimental demonstration of Airy beam PLS was reported. \\
\indent Importantly, as shown in our previous work \cite{Droulias2025}, bending beams can reach the same target user via multiple trajectories. This intriguing property naturally raises the question of how the PLS performance depends on the chosen trajectory and, importantly, whether there is an optimum trajectory that can maximize PLS. In this work, we analytically derive the various trajectories offered by such beams for the same user, and we perform an extensive study of their PLS performance. We analyze the dependencies between the possible locations of eavesdroppers and the bending beam design parameters, such as the beam curvature and the size of the transmitter antenna, in both line-of-sight (LoS) and non-LoS scenarios. We introduce metrics to assess their PLS performance and we demonstrate their superiority with respect to beams generated with conventional beam-forming.

\section{Theoretical analysis}
\subsection{System model}
\noindent We consider a communication link scenario, in which the transmitter (Tx) is equipped with a uniform linear array (ULA) and the receiver (Rx) with a single-antenna. The ULA is located at the origin of the coordinate system and is oriented along the $x$-axis, generating beams that propagate on the $xz$-plane, as shown in Fig.\,\ref{fig:fig01}. The beam profile at the input plane ($z=0$) is a function of $x$ and has the general form
\begin{align}
    E(x)=A(x) e^{j\phi(x)},
    \label{Eq:EqINPUTWAVE}
\end{align}
where $A$ is the amplitude and $\phi$ the phase. The beam profile \eqref{Eq:EqINPUTWAVE} takes discrete values at the locations of the ULA elements 
\begin{align}
    x_n=\left(n-\frac{N_x+1}{2} \right)d_x,
    \label{Eq:EqNx}
\end{align}
where $d_x$ is the inter-element spacing, and $n=1,2,\dots,N_x$.
The size of the ULA is thus $L_x=N_x d_x$. In what follows, we will keep the coordinate $x$ as a continuous variable to derive our analytical results, and in the simulations we will apply our analytical formulas at the discrete locations $x_n$.

\subsection{Trajectory design}

\begin{figure}
\centering
\includegraphics[width=\linewidth]{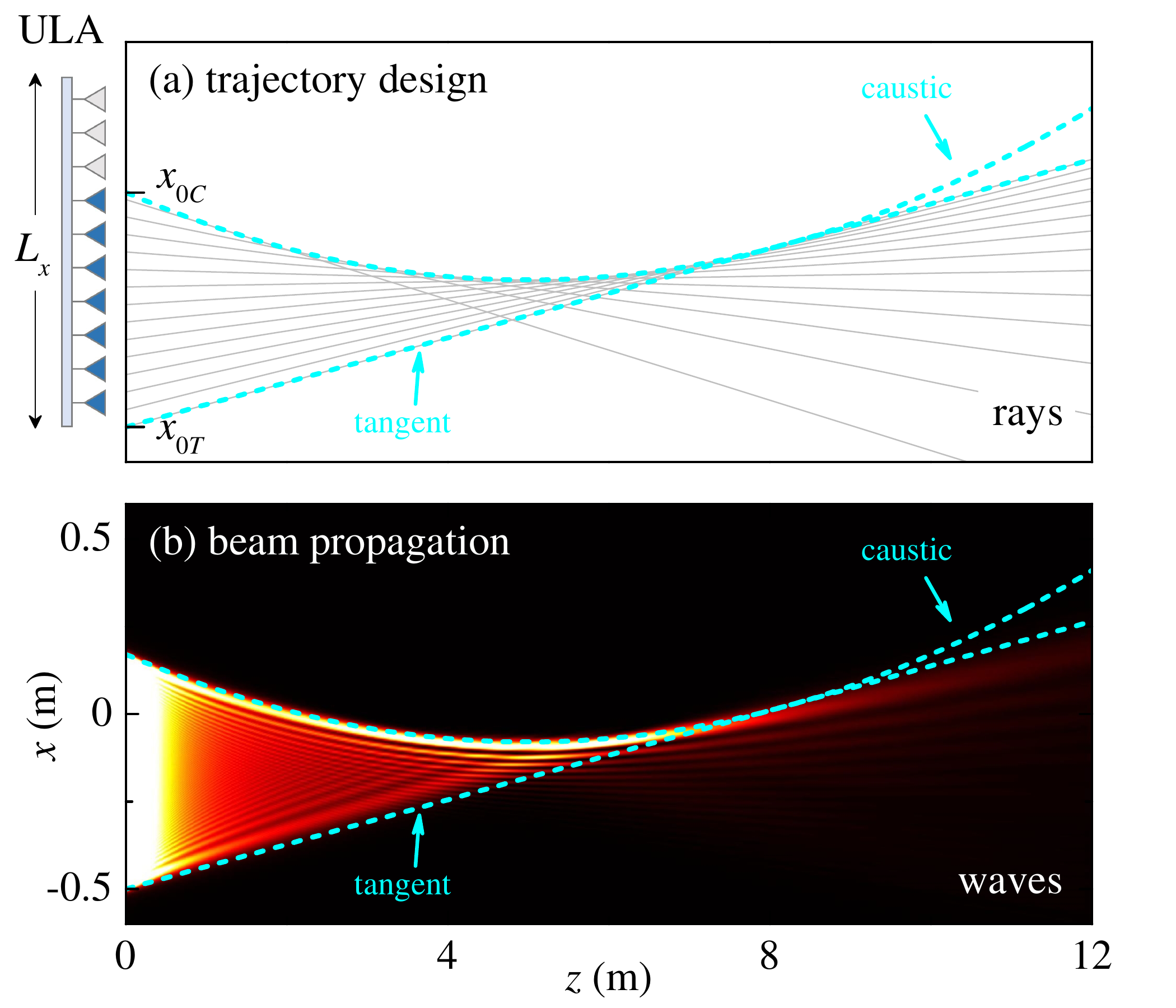}
\caption{Near-field engineering of bending beams, for a communication link between a Tx equipped with a ULA, and a single-antenna Rx. (a) Ray representation and (b) numerical propagation of equivalent beam. All rays in (a) are tangential to the caustic, which represents the trajectory of the beam’s main lobe shown in (b). The caustic and outer tangent are marked with cyan dashed lines. The ULA elements shown in the schematic in gray color are switched off.
}
\label{fig:fig01}
\end{figure}

\noindent To design beams that bend, we need to determine the necessary amplitude and phase of the ULA elements, so that the generated beam propagates along the prescribed bent path. Because the form of the trajectory is determined primarily by the input phase $\phi$ \cite{Droulias2025}, the amplitude $A$ across all ULA elements is taken as constant, significantly simplifying the excitation conditions in realistic systems. In this work, we consider the parabolic trajectory ($x_c,z_c$) with vertex located at ($x_0,z_0$), i.e.
\begin{align}
    x_c=x_0+\beta (z_c-z_0)^2,
    \label{Eq:EqCAUSTICparabolic}
\end{align}
where the subscript $c$ stands for \textit{caustic}. The caustic is the envelope of all the rays that form the bending beam (in the geometric optics approach), as illustrated in Fig.\,\ref{fig:fig01}(a); it corresponds to the trajectory of the beam's main lobe (in the wave approach), which is shown in Fig.\,\ref{fig:fig01}(b) for the equivalent beam. The caustic intersects the ULA at $x_{0C}=x_0+\beta z_0^2$, while the outer ray intersects the ULA at $x_{0T}=-L_x/2$, as indicated in Fig.\,\ref{fig:fig01}(a). \\
%
%
%
%
\indent To form the desired trajectory, i.e. caustic, one needs to engineer the slope of the rays at their origin, which is associated with the wavefront curvature of the wave, i.e. the phase of the beam at $z=0$ \cite{Droulias2025}. The application of this technique for the parabolic trajectory \eqref{Eq:EqCAUSTICparabolic}, leads to the input phase profile
\begin{align}    
    \phi(x) = 2\beta k z_0 x + \frac{4}{3} \beta^2 k \left(z_0^2 + \frac{x_0-x}{\beta} \right)^\frac{3}{2},
    \label{Eq:EqAiryPHASE}
\end{align}
which is real-valued for $x<x_0+\beta z_0^2\equiv x_{0C}$. Hence, excitation with elements only within $x_{0T}<x<x_{0C}$, as demonstrated in Fig.\,\ref{fig:fig01}(b), ensures that $\phi$ is purely real-valued.
\subsection{Propagation model}
\indent To simulate the propagation of the beam, we use the discrete model described in  \cite{Droulias2024c}, in which $E_\mathrm{obs}$, the $E$-field at the observation point $(x_\mathrm{obs},z_\mathrm{obs})$, is calculated numerically as 
\begin{equation}
    E_\mathrm{obs} = \sum_{n=1}^{N_x} E_n,
    \label{Eq:EqEr}
\end{equation}
where $E_n$ is the $E$-field from the $n^{th}$ ULA element to the observation point, which is at distance $r_n = \sqrt{(x_\mathrm{obs}-x_n)^2+z_\mathrm{obs}^2}$. The $E_n$-field is given by
\begin{equation}
    E_n= \sqrt{2Z_0 G_n U_n \frac{P_n}{4\pi}} \frac{e^{-j k r_n}}{r_n} e^{j\phi(x_n)}, 
    \label{Eq:EqEn}    
\end{equation}
where $Z_0$ is the free-space wave impedance, $G_n$ is the element gain, $U_n$ the element radiation pattern, and $P_n=\iint|E_n|^2/2Z_0$ the power radiated per element (integration performed on a surface enclosing the $n^{th}$ element). The steering vector that imposes the phase profile \eqref{Eq:EqAiryPHASE} on the elements of the ULA, is given by
\begin{align}
    \textbf{a}= \frac{1}{\sqrt{N_x}}\left[e^{j\phi(x_1)},e^{j\phi(x_2)},\dots, e^{j\phi(x_{Nx})} \right].
    \label{Eq:EqSV}
\end{align}
\indent In the example of Fig.\,\ref{fig:fig01}, the ULA has $N_x=1000$ elements and operates at $150\,\mathrm{GHz}$ with $d_x=\lambda/2=1\,\mathrm{mm}$, and hence $x_{0T}=-L_x/2=-0.5\,\mathrm{m}$. For simplicity, the ULA elements are omnidirectional ($G_n=U_n=1$). To form the trajectory shown in Fig.\,\ref{fig:fig01}(a), we have used the parameters $\beta = 0.01\,\mathrm{m^{-1}}$, $x_0 = -0.08\,\mathrm{m}$, $z_0 = 5\,\mathrm{m}$, for which $x_{0C}=0.17\,\mathrm{m}$. The input beam profile \eqref{Eq:EqAiryPHASE} is inserted in \eqref{Eq:EqEn}, and the field of each ULA element at the observation point is calculated with $P_n=1\,\mathrm{mW}$ for $x_{0T}<x_n<x_{0C}$ and $P_n=0\,\mathrm{mW}$ for $x_n>x_{0C}$, i.e. only a subset of the ULA elements is used. Fig.\,\ref{fig:fig01}(b) shows the calculated power density $|E_\mathrm{obs}|^2/2Z_0$ within a large area of interest in front of the ULA. \\

\subsection{Multi-trajectory design}
\noindent By choosing $\beta, x_0$ and $z_0$ as input parameters, the phase shifts imposed by \eqref{Eq:EqAiryPHASE} guarantee that the bending beam’s main lobe will evolve on the caustic trajectory $x_c=f(z_c)$ given by \eqref{Eq:EqCAUSTICparabolic}. Hence, the beam will reach any user at a location with coordinates $(x_\mathrm{Rx}, z_\mathrm{Rx})$ that satisfies condition $x_\mathrm{Rx} = f(z_\mathrm{Rx})$. 
Inversely, the location of the user can be chosen as an input parameter in \eqref{Eq:EqCAUSTICparabolic}, to determine the necessary $\beta, x_0$ and $z_0$ for the bending beam to reach the specific user. To ensure this, the caustic \eqref{Eq:EqCAUSTICparabolic} must be satisfied for $x_c = x_\mathrm{Rx}, z_c = z_\mathrm{Rx}$. Taking into account that the beam trajectory at $z_c=0$ crosses the $x$-axis at $x_c=x_{0C}$, we end up with two equations, $x_\mathrm{Rx} =x_0 + \beta(z_\mathrm{Rx}-z_0)^2$, $x_{0C} =x_0 + \beta(0-z_0)^2$, which are solved in terms of $x_0,z_0$, to provide the vertex of the parabola as
\begin{subequations}
    \begin{align}
     \label{Eq:EqX0}
        x_0 &= \frac{2 x_{0C} \left(x_\mathrm{Rx} + z_\mathrm{Rx}^2\beta \right) - x_{0C}^2 - \left(x_\mathrm{Rx} - z_\mathrm{Rx}^2 \beta \right)^2}{4 z_\mathrm{Rx}^2 \beta},\\ 
        \label{Eq:EqZ0}
        z_0 &= \frac{z_\mathrm{Rx}^2 \beta + x_{0C} - x_\mathrm{Rx}}{2 z_\mathrm{Rx} \beta}.
    \end{align}
\end{subequations}
Hence, a user at a certain location $(x_\mathrm{Rx}, z_\mathrm{Rx})$ can be served by multiple bending beams, determined by the choice of parameters $x_{0C}$ and $\beta$.
%

%
\begin{figure}
\centering
\includegraphics[width=\linewidth]{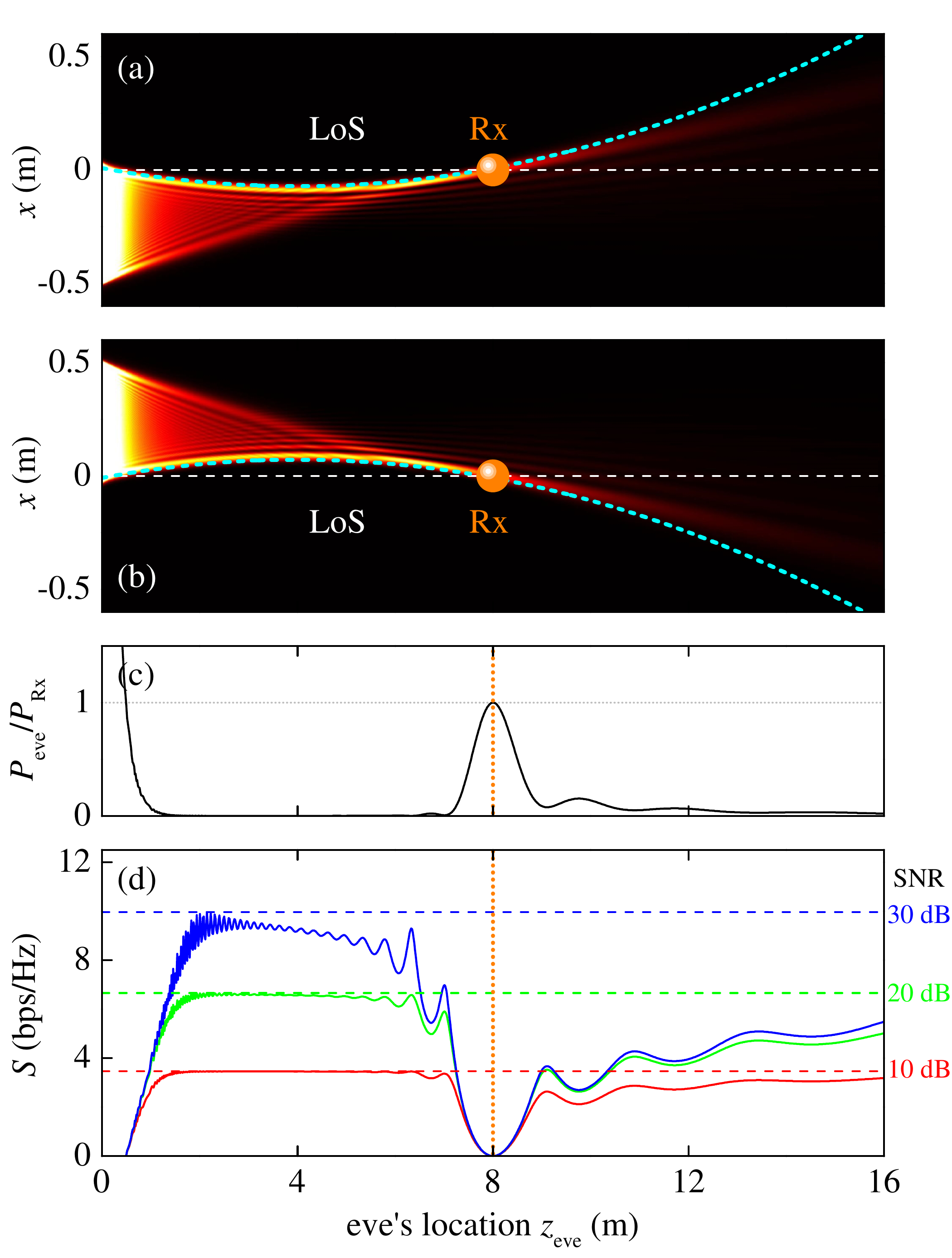}
\caption{Performance of bending beams for LoS eavesdropping. (a) Beam with $\beta=5\times10^{-3} \,\mathrm{m^{-1}}$, $x_{0T}=-0.5\,\mathrm{m}$, $x_{0C}=0\,\mathrm{m}$. (b) Beam with $\beta=-5\times10^{-3}\,\mathrm{m^{-1}}$, $x_{0T}=0.5\,\mathrm{m}$, $x_{0C}=0\,\mathrm{m}$. (c) Power ratio $P_\mathrm{eve}/P_\mathrm{Rx}$, as a function of the eve's location along the Tx-Rx LoS, for a user located at $(x_\mathrm{Rx}, z_\mathrm{Rx})=(0\,\mathrm{m},8\,\mathrm{m})$. (d) Secrecy rate, as a function of the eve's location along the Tx-Rx LoS, for different SNR levels. The dashed lines mark the theoretically maximum achievable $S$ for each SNR.}
\label{fig:fig02}
\end{figure}
%

%
\begin{figure}
\centering
\includegraphics[width=\linewidth]{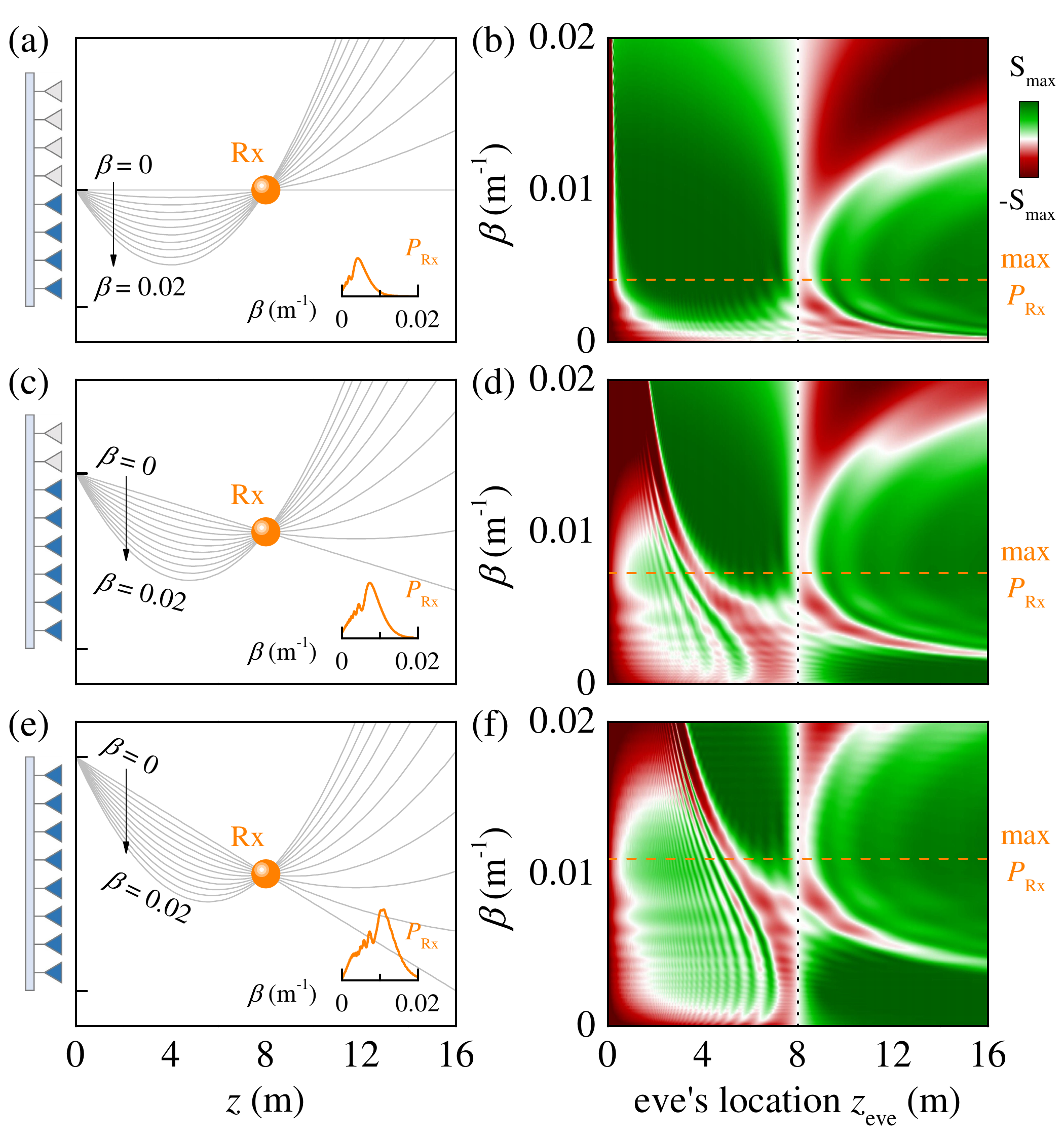}
\caption{Performance of multi-trajectory bending beam design for LoS eavesdropping. Bending beam excitation with $x_{0T}=-0.5\,\mathrm{m}$, and (a),(b) $x_{0C}=0\,\mathrm{m}$, (c),(d) $x_{0C}=0.25\,\mathrm{m}$, and (e),(f) $x_{0C}=0.5\,\mathrm{m}$. (a),(c),(e) Beam trajectories, as a function of the beam's curvature; the inset shows the power at the Rx, as the parameter $\beta$ varies. (b),(d),(f) Secrecy rate, as a function of the eve's location along the Tx-Rx LoS, and the curvature parameter $\beta$. The vertical dotted line marks the Rx location, and the horizontal dashed line marks the maximum power at the Rx for each case. The ULA elements shown in the schematic in gray color are switched off.}
\label{fig:fig03}
\end{figure}
%
%
\section{PLS performance assessment}
\subsection{LoS eavesdropping}
\noindent In our eavesdropping scenario, the user is located at $(x_\mathrm{Rx}, z_\mathrm{Rx})=(0\,\mathrm{m},8\,\mathrm{m})$ and the eve at $(x_\mathrm{eve}=0\,\mathrm{m}, z_\mathrm{eve})$, i.e. moves along the Tx-Rx LoS, as illustrated in Fig.\,\ref{fig:fig02}(a). The Tx sends a bending beam towards the Rx with parameters $\beta=5\times10^{-3} \,\mathrm{m^{-1}}$, $x_{0T}=-0.5\,\mathrm{m}$, $x_{0C}=0\,\mathrm{m}$ (half the ULA elements are utilized), practically circumventing the eve. An opposite sign in $\beta$ produces a mirror-symmetric beam, as shown in Fig.\,\ref{fig:fig02}(b). The ratio of the power captured by the eve, $P_\mathrm{eve}$, over the power received by the Rx, $P_\mathrm{Rx}$ is shown in Fig.\,\ref{fig:fig02}(c), and is identical for both beams. As expected, the power at the eve is significantly suppressed in the Tx-Rx LoS because of beam bending, except perhaps very close to the Tx or the Rx. To assess the PLS security in this scenario, we calculate the secrecy rate $S$, which is the difference between the capacity of the legitimate user (the Rx in our case) and the eavesdropper \cite{Barros2006,DrouliasPLS}, i.e.
\begin{align}
        S = \log_2(1+\mathrm{SNR_\mathrm{Rx}})-\log_2(1+\mathrm{SNR_\mathrm{eve}}),
\label{Eq:EqSR}
\end{align}
where $\mathrm{SNR_\mathrm{eve}}=P_\mathrm{eve}/P_N$ is the SNR (signal-to-noise ratio) at the eve, $\mathrm{SNR_\mathrm{Rx}}=P_\mathrm{Rx}/P_N$ is the SNR at the Rx, and $P_N$ is the noise power. To ensure secure communications $S$ must be as high as possible, whereas low or even negative values indicate that the link security is severely compromised. In Fig.\,\ref{fig:fig02}(d) we present the secrecy rate $S$ of the bending beams considered in our eavesdropping scenario (identical for both beams), for three SNR levels. The dashed lines mark the maximum $S$ for each noise level, which is the limit of \eqref{Eq:EqSR} for $P_\mathrm{eve}\rightarrow0$. In essence, when the eve is at the Tx-Rx LoS, the bending beam entirely protects the user from eavesdropping. \\
\indent In Fig.\,\ref{fig:fig03} we extend the study of our eavesdropping scenario to different bending beam designs, all tailored to reach the Rx through different paths. In Fig.\,\ref{fig:fig03}(a) we again use half the ULA elements ($x_{0C}=0\,\mathrm{m}$) and tune the beam curvature within the range $\beta (\mathrm{m^{-1}})\in[0,0.02]$. Using \eqref{Eq:EqX0} and \eqref{Eq:EqZ0} we calculate the parameters $x_0,z_0$ for each trajectory, which we use in \eqref{Eq:EqAiryPHASE} to generate the corresponding beam. The inset shows the power received by the Rx, which changes according to how efficiently the beam reaches the Rx, and is maximized for $\beta=0.0041\,\mathrm{m}^{-1}$. In Fig.\,\ref{fig:fig03}(b) we present the calculated secrecy rate $S$ along the Tx-Rx LoS, as a function of the beam's curvature. The secure regions $(S>0)$ are shown in green, while the vulnerable regions $(S<0)$ are shown in red. We observe that there is a wide range of $\beta's$ that provides high $S$ for eve located either in front or behind the Rx, with the optimum choice corresponding to the $\beta$ that maximizes the power at the Rx. In Fig.\,\ref{fig:fig03}(c),(d) we repeat the calculations for beams generated using $75\%$ of the ULA elements ($x_{0C}=0.25\,\mathrm{m}$), and in Fig.\,\ref{fig:fig03}(e),(f) using all ULA elements ($x_{0C}=0.5\,\mathrm{m}$). Using a larger area of the ULA results in non-negligible fields in front of the Tx, which is why the secrecy rate reduces when the eve resides between the Tx-Rx, and is restored with higher $\beta's$ that produce stronger bending, thus suppressing the fields in front of the Tx.

%
\begin{figure}
\centering
\includegraphics[width=\linewidth]{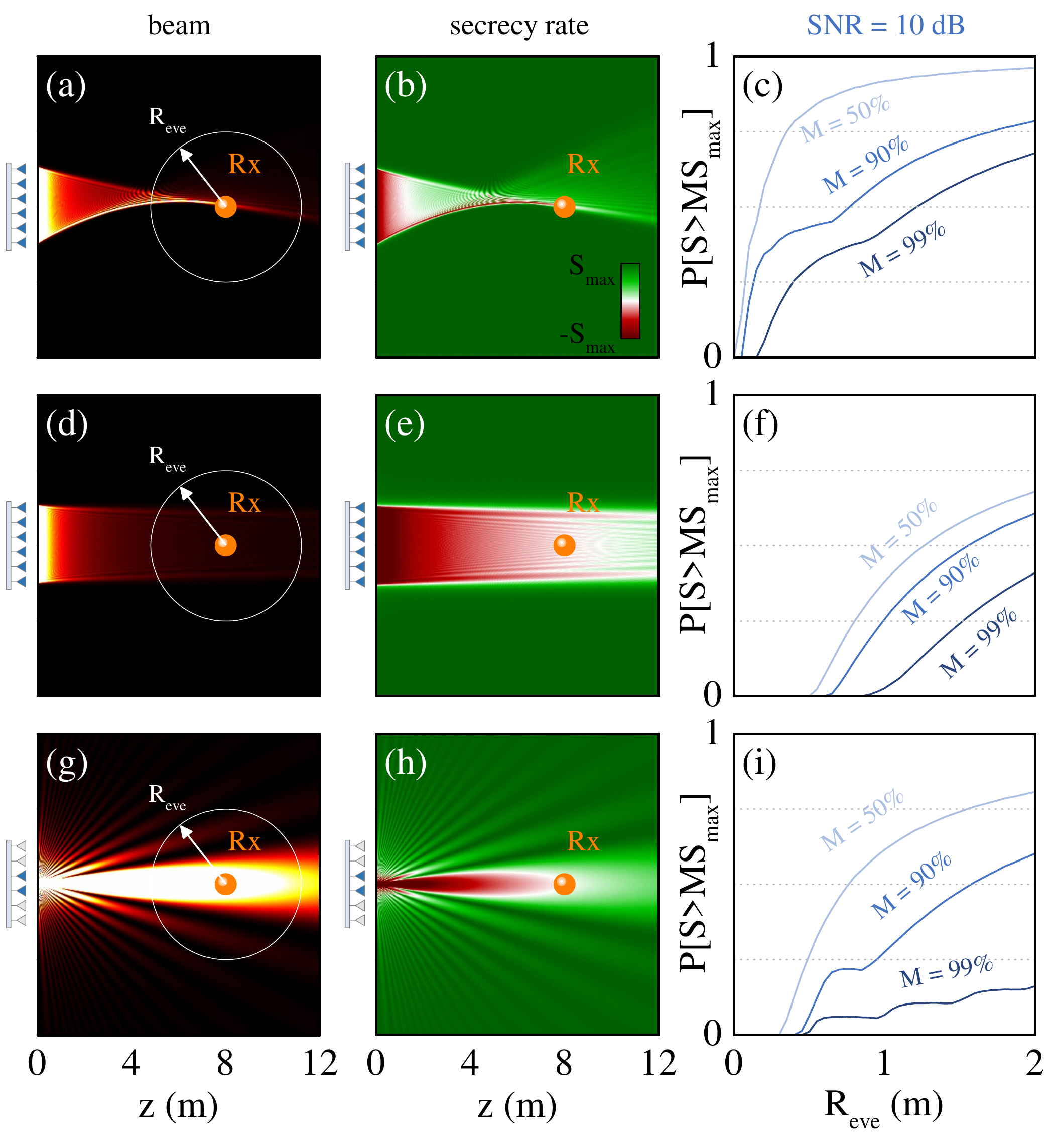}
\caption{PLS performance beyond the LoS. (a),(b),(c) Bending beam with $\beta=0.015\,\mathrm{m}^{-1}$, (d),(e),(f) beam forming, and (g),(h),(i) beam forming using only $N_x=32$ elements. (a),(d),(g) Propagated beams, also showing the outline of the disk, in which the eve is located. (b),(e),(h) Secrecy rate for $\mathrm{SNR=10\,dB}$. (c),(f),(i) Probability of $S>MS_\mathrm{max}$ for different values of $M$, as a function of the disk radius within which the eve is located.}
\label{fig:fig04}
\end{figure}
%
\subsection{Eavesdropping beyond the LoS}
\noindent In Fig.\,\ref{fig:fig04} we study the PLS for eve located within a disk of radius $R_\mathrm{eve}$ around the Rx, i.e. not limited to the Tx-Rx LoS; the Rx location is the same as in the previous examples and the eve is located anywhere within the disk with uniform probability. The Tx utilizes the entire ULA to generate a bending beam with $\beta=0.015\,\mathrm{m}^{-1}$, as shown in Fig.\,\ref{fig:fig04}(a). 
In Fig.\,\ref{fig:fig04}(b) we present the calculated secrecy rate for $\mathrm{SNR}=10\,\mathrm{dB}$ for all possible eve locations within a large disk that covers the entire shown area. 
Because the spatial distribution of $S$ is determined by the beam's spatial power distribution relative to its power at the Rx location, $S$ essentially imprints the beam shape in space, providing a map of secure (green) and vulnerable areas (red).
To assess the bending beam PLS performance, in Fig.\,\ref{fig:fig04}(c) we calculate the probability of the ratio $S/S_\mathrm{max}$ being larger than a threshold value $M$, and we present our results for $M=50\%,90\%,99\%$, as a function of the disk radius $R_\mathrm{eve}$. To calculate this probability, for each $R_\mathrm{eve}$ we choose several random locations of the eve within the disk, and count the number of locations with $S>MS_\mathrm{max}$ over all possible eve's locations within the chosen disk.
%
\begin{figure}
\centering
\includegraphics[width=\linewidth]{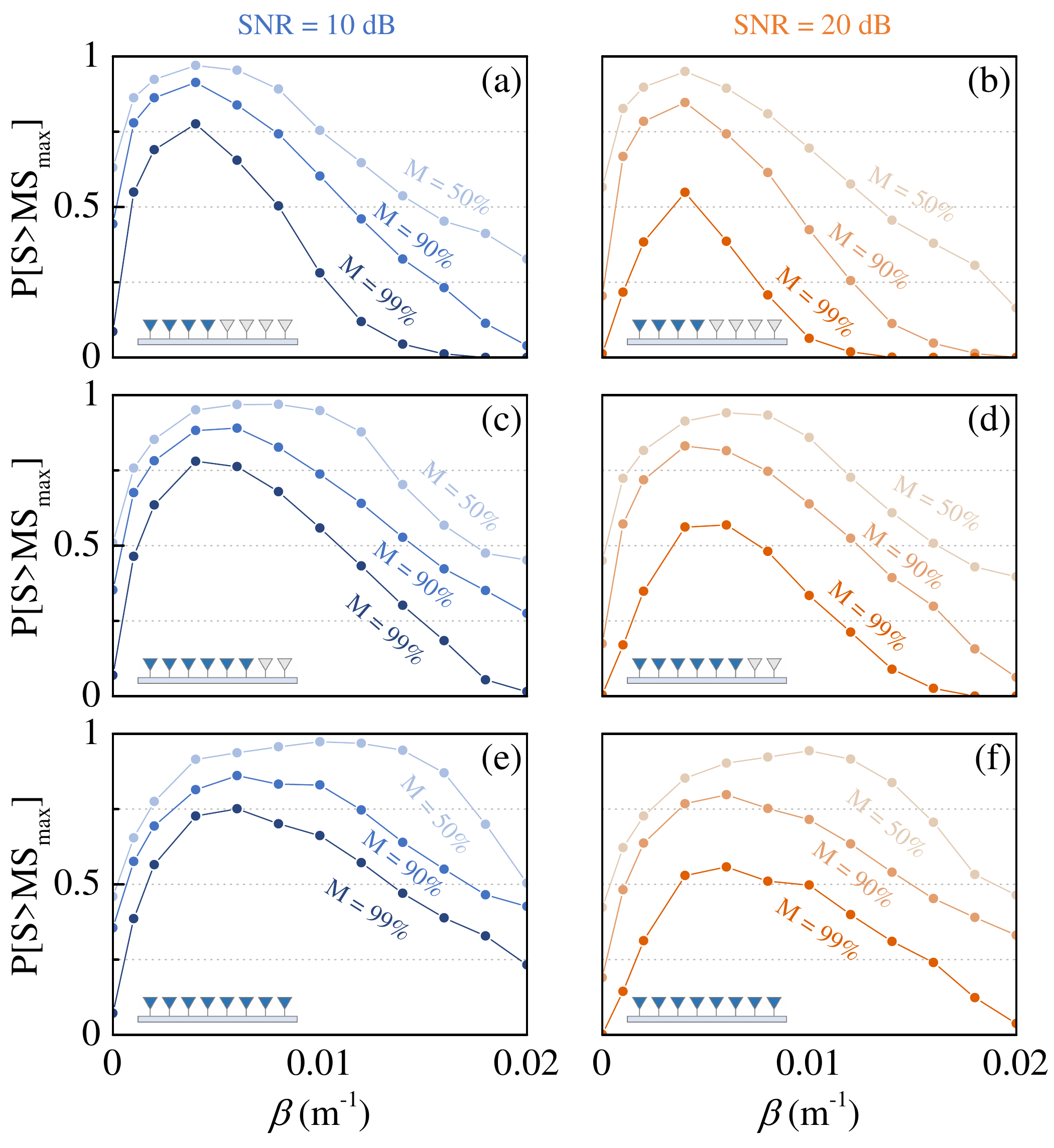}
\caption{Bending beam PLS efficiency, as a function of the trajectory curvature parameter $\beta$. Beam excitation with $x_{0T}=-0.5\,\mathrm{m}$, and (a),(b) $x_{0C}=0\,\mathrm{m}$, (c),(d) $x_{0C}=0.25\,\mathrm{m}$, and (e),(f) $x_{0C}=0.5\,\mathrm{m}$. (a),(c),(e) $\mathrm{SNR=10\,dB}$. (b),(d),(f) $\mathrm{SNR=20\,dB}$. The ULA elements shown in the schematic in gray color are switched off.}
\label{fig:fig05}
\end{figure}
%
To compare our results with conventional beamforming, in Figs.\,\ref{fig:fig04}(d),(e),(f) we repeat the calculations using \eqref{Eq:EqAiryPHASE} with $\phi=0$, to eliminate the input wavefront curvature. Due to the wide extent of the beam, the resulting secure areas are now limited with respect to those of the bending beam. This is not unexpected because, for the first few meters of propagation, the beam propagates within its near field with beamwidth comparable to the ULA extent (the Fraunhofer distance in this case is $\sim 1\,\mathrm{km}$). Importantly, simple inspection of Fig.\,\ref{fig:fig04}(c) and Fig.\,\ref{fig:fig04}(f) reveals that bending beams outperform conventional beamforming in terms of their PLS capabilities.
Last, to compare our findings with conventional far-field beamforming, in Figs.\,\ref{fig:fig04}(g),(h),(i) we use only 32 ULA elements, to generate a beam that quickly enters the Fraunhofer zone (the Fraunhofer distance in this case is $\sim 1\,\mathrm{m}$). We observe that, due to diffraction, the beam again extends over a wide area, leading to limited secure areas, which in turn lead to poor PLS performance. \\
\indent The unique property of bending beams to offer multiple trajectories for the same Rx naturally raises the question of whether there is an optimum trajectory that maximizes PLS. To address this aspect, in Fig.\,\ref{fig:fig05} we study the PLS performance with respect to the bending beam design parameters. We assume that the eve is located somewhere within a disk of radius $R_\mathrm{eve}=1\,\mathrm{m}$ with uniform probability, and in Fig.\,\ref{fig:fig05} we extend the calculations of Fig.\,\ref{fig:fig04}(c) for a multitude of $\beta's$.
First, we use half the ULA elements ($x_{0C}=0\,\mathrm{m}$) and calculate the probability of $S>MS_\mathrm{max}$, which serves as the PLS efficiency. In Fig.\,\ref{fig:fig05}(a),(b) we present the results for $\mathrm{SNR=10\,dB,20\,dB}$ and different values of the parameter $M$, which reveal that there is indeed an optimum $\beta$.
In Fig.\,\ref{fig:fig05}(c),(d) we repeat the calculations for beams generated using $75\%$ of the ULA elements ($x_{0C}=0.25\,\mathrm{m}$), and in Fig.\,\ref{fig:fig05}(e),(f) using all ULA elements ($x_{0C}=0.5\,\mathrm{m}$). 
We observe that, with larger ULA aperture, there is a wide range of $\beta's$ that provides similar levels of coverage, which gradually becomes limited with the requirement for higher $S$. This property is crucial for PLS, as it enables the usage of multiple bending beams with similar performance to suppress eavesdropping.

\section{Conclusion}

\noindent In this work we studied the PLS performance of bending beams. We analytically determined the various trajectories offered by such beams for the same user, and we analyzed the dependencies between the possible locations of eavesdroppers and the bending beam design parameters, such as the beam curvature and the fraction of utilized ULA elements. We introduced metrics to assess their PLS performance and we demonstrated their superiority with respect to beams generated with conventional beam-forming in both LoS and non-LoS scenarios.

\section*{Acknowledgment}

This work was supported by the European Commission’s Horizon Europe Programme under the Smart Networks and Services Joint Undertaking INSTINCT project (Grant Agreement $101139161$).

\bibliographystyle{IEEEtran}
\bibliography{IEEEabrv,main}

\end{document}